\documentclass[12pt]{article}
\usepackage{epsf}
\usepackage{amsmath}
\usepackage{graphics}
\usepackage{cite}

\renewcommand{\thefootnote}{\#\arabic{footnote}}
\begin{document}

\newcommand{\gtrsim}{ \mathop{}_{\textstyle \sim}^{\textstyle >} }
\newcommand{\lesssim}{ \mathop{}_{\textstyle \sim}^{\textstyle <} }

\newcommand{\rem}[1]{{\bf #1}}

\renewcommand{\thefootnote}{\fnsymbol{footnote}}
\setcounter{footnote}{0}
\begin{titlepage}

\def\thefootnote{\fnsymbol{footnote}}

\begin{center}
\hfill August 2015\\
\vskip .5in
\bigskip
\bigskip
{\Large \bf Scale Invariant Density Perturbations from Cyclic Cosmology}

\vskip .45in

{\bf Paul Howard Frampton}

{\em Oxford, UK.}

\end{center}

\vskip .4in
\begin{abstract}
It is shown how quantum fluctuations of the radiation during the contraction era
of a CBE (Comes Back Empty) cyclic cosmology can provide density fluctuations which
re-enter the horizon during the subsequent expansion era and at lowest order are scale
invariant, in a Harrison-Zel'dovich-Peebles sense, as necessary to
be consistent with observations of large scale structure.
\end{abstract}
\end{titlepage}

\renewcommand{\thepage}{\arabic{page}}
\setcounter{page}{1}
\renewcommand{\thefootnote}{\#\arabic{footnote}}

\newpage

\section{Introduction}
In a cyclic cosmology based on the CBE (Comes back empty) assumption
\cite{PHF1,PHF2} it has been shown that flatness ($\Omega \simeq 1$) is
achieved with high precision without an inflationary era. It arises from two reasons,
firstly that during the contraction phase $\Omega=1$ is approached naturally
by the Friedmann equation and secondly the significant reduction in size
of the contracting introverse relative to the expanding extroverse makes the
flatness even more precise.

\bigskip

\noindent
The CBE cyclic universe is motivated by satisfying the second law of thermodynamics
where the propensity of entropy to increase provides a well-known stumbling
block \cite{Tolman}. By the fact that general relativity is valid for 99\% of the cycle,
and joining the radiation-dominated contraction and early expansion eras, the
duration of the expansion and contraction eras were estimated \cite{PHF2} as $1.3$ trillion
years, close to a hundred times the present age.

\bigskip

\noindent
Another requirement of a cosmological theory is to explain the observed density 
fluctuations which set up the initial conditions for structure formation. The necessary
properties can be deduced from the anisotropy of the Cosmic Microwave Background (CMB),
also from the distribution of matter as revealed by galaxy redshift surveys.

\bigskip

\noindent
The density fluctuation $\delta({\bf x})$ is defined by
\begin{equation}
\delta({\bf x}) \equiv \frac{\rho({\bf x})}{\bar{\rho}} - 1 = 
\int dk ~\delta_{\bf k} ~ \exp(i {\bf k.x})
\label{fluctuation}
\end{equation}
where $\bar{\rho}$ is the average density and ${\bf k}$ is the wave number of the fluctuation.

\bigskip

\noindent
The isotropic power spectrum $P(k)$ with $k \equiv |{\bf k}|$ is defined by the two-point function
\begin{equation}
< \delta_{\bf k} \delta_{\bf k^{'}} > = \frac{2 \pi^2}{k^3} \delta({\bf k} - {\bf k^{'}}) P(k)
\label{power}
\end{equation}
and the behavior of $P(k)$ is characterized by
\begin{equation}
P(k) \propto k^{n_s - 1}
\label{scaling}
\end{equation}
where $n_s$ is the scalar spectral index. $n_s=1$ corresponds to scale invariance
\cite{Harrison, Zeldovich,Peebles}.
The latest determination of $n_s$ from Planck \cite{Planck1,Planck2} gives
\begin{equation}
n_s = 0.9655 \pm 0.0062
\label{index}
\end{equation}
at $68\%$ confidence level so there is approximate scale invariance. The fact that $n_s<1$
implies a spectral tilt towards the red because lower frequencies are enhanced relative
to the exactly scale-invariant $n_s=1$ case. In the present article, we shall be content to
derive scale-invariant fluctuations with $n_s=1$ and leave the study of the higher order
corrections which redden the spectrum to $n_s\simeq0.96$ for future research.

\bigskip

The plan of this paper is that in Section 2 the CBE cyclic cosmological model
is disussed, in Section 3 density perturbations in inflationary cosmology are reviewed
as a preparation, in Section 4 the density perturbations in CBE cyclic cosmology are
derived, and finally in Section 5 there is discussion.

\newpage

\section{CBE Model}

\noindent
The CBE cosmological model makes use of the superluminal expansion
first observed in 1998 \cite{Perlmutter,Riess}. It separates the universe
into the visible universe, or particle horizon, in CBE called the introverse, and 
the remaining part of spacetime outside the introverse in CBE called
the extroverse.

\bigskip

\noindent
The motivation for the CBE assumption is consistency with the second law
of thermodynamics and explanation of the vanishing entropy at the
beginning of expansion.  

\bigskip

\noindent
The idea first introduced in 2007 \cite{BF} is that only the introverse 
be considered because everything outside is unobservable, and further
that the introverse be chosen, as is true for almost every choice, to be
empty of matter and to include only dark energy together with tiny
amounts of radiation and curvature.

\bigskip

\noindent
One striking property of the observed geometry of the visible universe
is its proximity to flatness $\Omega_{TOTAL} = 1$.

\bigskip

\noindent
In \cite{PHF2}, the turnaround time was established as $t_T = 1.3$Ty at which 
time the radii of the introverse(IV) and extroverse(EV) are respectively
$R_{IV}(t_T) = 58$Gly and $R_{EV}(t_T) = 4.4\times10^{42}$Gly. This implies
in the notation of \cite{PHF1} that 
\begin{equation}
f(t_T)=R_{IV}(t_T)/R_{EV}(t_T) = 1.1 \times 10^{-41}
\label{tT}
\end{equation}
which can be substituted in
\begin{equation}
|\Omega(t_B) - 1| = f(t_T)^4 C_{\Omega} t_B
\end{equation}
in which $C_{\omega}= 3.9 \times 10^{-16} s^{-1}$ and $t_B$ is the bounce time
given by
\begin{equation}
t_B =  10^{-44}s  \left( \frac{10^{19} GeV}{T_B} \right)^2
\end{equation}
where $T_B$ is the bounce temperature in GeV. Taking typical temperature values
$T_B = 10^6, 10^{11}, 10^{16}$ GeV the bounce times are
$t_B = 10^{-18}, 10^{-28}, 10^{-38}$ s.

\bigskip

\noindent
However, because $f(t_T)$ in Eq.(\ref{tT}) is so extremely small, when we
calculate the value of the present total density for any of these $t_B$,
the result for $|\Omega_{TOTAL}(t_0) -1|$ is infinitesimal, well below
an inverse googol, and its exact result become academic:
\begin{equation}
|\Omega_{TOTAL}(t_0) - 1| \ll 10^{-100}
\end{equation}
which is interesting.

\bigskip

\noindent
This implies that any departure from $\Omega_{TOTAL}(t_0) =1$
can falsify the CBE model. The expected limit to the observational
accuracy in the measurement is comparable to the size of the
observed perturbations $\sim 10^{-5}$. Therefore falsification
of CBE would follow from
\begin{equation}
|\Omega_{TOTAL}(t_0)-1| > 10^{-5}
\label{test}
\end{equation}

\bigskip

\noindent
On the other hand, the smaller this quantity becomes, the more it will
favor CBE over inflation which predicts $|\Omega(t_0)-1|$ to be small
but not identically zero.

\bigskip

\noindent
The contraction period is radiation dominated and begins
at the turnaround $t_T\sim1.3$ Ty. The introverse at that time has a scale factor
$\hat{a}(t_T) = 1.11$  and subsequently contracts as $\hat{a}(\hat{t}) \sim \hat{t}^{1/2}$
where for convenience during contraction we define a displaced time $\hat{t}$ by
$\hat{t} \equiv (t_B-t)$ with $t_B$ the
time of the bounce, which is $t_B\sim 2.6$Ty.  

\bigskip

\noindent
In terms of $\hat{t}$ the contraction scale
factor $\hat{a}(\hat{t})$ therefore shrinks according to
\begin{equation}
\hat{a} (\hat{t})  = \hat{a}(t_T) \left( \frac{\hat{t}}{t_T} \right)^{\frac{1}{2}}
\label{ahat}
\end{equation}
for $0 < \hat{t} < t_T$ and must  be matched on to the expansion scale factor $a(t)$ at the onset its matter domination
$t=t_m$ so that 
\begin{equation}
\hat{a}(\hat{t}=t_m) = a(t_m) = 2.1\times 10^{-4}
\label{matching}
\end{equation}
and from then to and from the bounce the expansion $a(t)$ and
the contraction $\hat{a}(t)$ are equal.

\bigskip

\noindent
So far the discussion is classical without fluctuations. Density fluctuations are
expected to arise from quantum effects so first we shall discuss how this happens
in inflationary models in the next section, then show how the CBE model can 
produce scale-invariant density perturbations.

\newpage

\section{Density perturbations with inflation}
In the inflationary scenario, the superluminal accelerated expansion during inflation
makes quantum fluctuations of the inflaton enlarge to macroscopic
size and freeze in after leaving the horizon. They later re-enter the horizon as
classical density perturbations which provide the initial conditions necessary for
structure formation. 

\bigskip

\noindent
Let us flesh out some of the mathematical details, extracted from \cite{Lyth}. This is
not original but will be useful in addressing the perturbation issue for the CBE model in
the subsequent section.

\bigskip

\noindent
We take a single inflaton field $\phi({\bf x}, t)$ in a locally flat spacetime as in \cite{Guth,Linde,Steinhardt} whereupon its classical field equation is
\begin{equation}
\ddot{\phi} + 3 H \dot{\phi} - a^{-2}\nabla^2\phi + V^{'} = 0
\label{cfe}
\end{equation}
with $V^{'} = dV/d\phi$.

\bigskip

\noindent
At lowest order, a perturbation $\delta\phi_k$ for wave number $k$ therefore satisfies

\begin{equation}
\ddot{\delta\phi}_k + 3 H \dot{\delta\phi}_k + \left( \frac{k}{a}\right)^2 \delta\phi_k
+ V^{''} \delta\phi_k = 0
\label{pertcfe}
\end{equation}
with $V^{''} = d^2V/d\phi_k^2$.

\bigskip

\noindent
For a light field $V \ll H$ and $V \ll (k/a)^2$, so that
\begin{equation}
\ddot{\delta\phi}_k + 3 H \dot{\delta\phi}_k + \left( \frac{k}{a}\right)^2 \delta\phi_k = 0
\label{pertcfelessV}
\end{equation}

\bigskip

\noindent
We are concerned only with a few Hubble times $H^{-1}$ around the exit
time during which we may take the slowly varying $H$ to be constant at
$H_k$. Defining conformal time $\eta$ by $\eta=-1/aH$, we find an oscillator
equation
\begin{equation}\frac{d^2\phi_k(\eta)}{d^2\eta^2} + \omega_k^2(\eta)\phi_k(\eta) = 0
\end{equation}
where
\begin{equation}
\omega_k^2 = k^2 - \left( \frac{2}{\eta^2} \right) = k^2 - 2(a H_k)^2
\label{omega}
\end{equation}

\bigskip

\noindent
Before horizon exit there is constant wave number $k$. In a small spacetime
region with $k^{-1} \ll \Delta\eta \ll (a H_k)^{-1}$ there are still many oscillations
but the spacetime curvature is negligible. During the interval $\Delta\eta$, $k$
is the physical wave number because the second term in Eq.(\ref{omega})
is negligible.

\bigskip

\noindent
We can decompose the Fourier component $\hat{\phi}_k(\eta)$ as
\begin{equation}
(2 \pi)^3 \hat{\phi}_k(\eta) = \phi_k(\eta) \hat{a}({\bf k}) + \phi_k^*(\eta)\hat{a}^{\dagger}(-{\bf k})
\label{kmode}
\end{equation}

\bigskip

\noindent
The required solution is
\begin{equation}
\phi_k(\eta) = \frac{e^{-ik\eta}}{\sqrt{2k}} \frac{(k\eta-i)}{k\eta}
\end{equation}
which long after the horizon exit approaches
\begin{equation}
\phi_k(\eta) = - \frac{i}{\sqrt{2k}} \frac{1}{k\eta}
\end{equation}

\bigskip

\noindent
It is assumed that the state corresponds to vacuum with no particles
and $<\phi_k>=0$.  We have a gaussian random field whose ensemble average
is the vacuum expectation value. The power spectrum is defined by the
two-point function
\begin{equation}
<\phi_{\bf k}\phi_{\bf k^{'}}> = \frac{2 \pi^2}{k^3} {\cal P}_{\phi}(k) \delta^3({\bf k} + {\bf k^{'}})
\end{equation}

\bigskip

\noindent
Using the earlier expressions, we now find the scale invariant result
\begin{equation}
{\cal P}_{\phi}(k) = \left( \frac{H_k}{2 \pi} \right)^2
\label{scaleinvariant}
\end{equation}
where $H_k$ is a constant, $H_k = H$.

\bigskip

\noindent
The mean square perturbation of the field is given by
\begin{equation}
<\delta\phi^2 ({\bf x}, t) >  = \left( \frac{H}{2 \pi} \right)^2 N(t)
\end{equation}
where $N(t)$ is the number of e-folding of inflation after leaving the horizon.

\bigskip

\noindent
Finally, long after the fluctuation leaves the horizon the k-mode in Eq.(\ref{kmode})
becomes purely imaginary with a definite constant value for measurements and
hence can be regarded as classical. These density perturbations re-enter the
horizon later in the expansion.

\bigskip

\noindent

\newpage

\section{Density perturbations in CBE model}
Because the photon is massless, the CBE contraction era is classically
scale invariant. There is no scalar inflaton and the quantum fluctuations of relevance are
in the electromagnetic field $A_{\mu}({\bf x}, t)$.

\bigskip

\noindent
In inflation, quantum fluctuations of the inflaton freeze-out, exit the horizon during
expansion then later in the expansion re-enter the horizon
as classical density perturbations.

\bigskip

\noindent
In the CBE cyclic scenario quantum fluctuations of the radiation field freeze in after 
leaving the horizon {\it during contraction} and later re-enter the horizon as
classical density perturbations {\it after the bounce} when the universe is expanding,
thereby providing the initial conditions necessary for structure formation. 

\bigskip

\noindent 
The classical field equation for a perturbation $\delta(A_{\mu})_k$ may be cast into the simple harmonic oscillator equation
\begin{equation}
\frac{d^2(A_{\mu})_k(\eta)}{d^2\eta^2} + \omega_k^2(\eta)(A_{\mu})_k(\eta) = 0
\end{equation}
where
\begin{equation}
\omega_k^2 = k^2 - \left( \frac{2}{\eta^2} \right) = k^2 - 2(a H_k)^2
\label{omegaA}
\end{equation}

\bigskip

\noindent
Quantum k-modes are then defined by
\begin{equation}
(2\pi)^3 (\hat{A}_{\mu})_k(\eta) = (A_{\mu})_k(\eta) \hat{a}(k) + (A_{\mu})_k^*(\eta) \hat{a}^{\dagger} (-k)
\end{equation}

\bigskip

\noindent
Subject to the initial condition
\begin{equation}
(A_{\mu})(\eta) = \frac{1}{\sqrt{2k}} \epsilon_{\mu}(k) e^{-ik\eta},
\end{equation}
the appropriate solution is then
\begin{equation}
(A_{\mu})_k(\eta)= \epsilon_{\mu}(k) \frac{e^{-ik\eta}}{\sqrt{2k}}  \frac{(k\eta-i)}{k\eta}
\end{equation}

\bigskip

\noindent
Well after exiting the horizon, this solution becomes purely imaginary
and freezes in with time-independent eigenvalues just as if classical, namely
\begin{equation}
(A_{\mu})_k(\eta) = - \epsilon_{\mu}(k)  \frac{i}{\sqrt{2k}} \frac{1}{k\eta}
\end{equation}

\bigskip
\bigskip

\noindent
The two point function may be defined in the 't Hooft-Feynman gauge by
\begin{equation}
<(A_{\mu})_{\bf k} (A_{\nu})_{{\bf k^{'}}}> = 
\frac{2\pi^2}{k^3} \left( g_{\mu\nu} - \frac{k_{\mu}k_{\nu}}{k^2} 
\right) {\cal P}_A(k) \delta^3({\bf k} + {\bf k^{'}})
\end{equation}

\bigskip

\noindent
Proceeding with the parallel steps as before one arrives at the scale-invariant
power law, at this lowest order
\begin{equation}
{\cal P}_A (k) = \left( \frac{H_k}{2\pi} \right)^2.
\end{equation}

\bigskip

\noindent
The mean square perturbation is given by considering a comoving box of side $(aL)$
by 
\begin{equation}
< \left| \delta A_{\mu}({\bf x},t) \right|^2> =
\left( \frac{H}{2\pi} \right)^2 \int_{L^{-1}}^{aH} \frac{dk}{k} = \left( \frac{H}{2\pi} \right)^2 \ln(LHa)
\end{equation}

\bigskip
\bigskip

\noindent
These are the scale invariant density perturbations in the CBE model. They do not
re-enter the horizon during the contraction era but only after the bounce. That is when
they enter as frozen-in classical density perturbations and seed large-scale structure
formation.

\bigskip

\noindent
There is no superluminal accelerated expansion in the early universe, only 
starting at $t\simeq 9.8$Gy when dark energy begins to dominate
over matter very much later in the expansion era.

\bigskip

\noindent
In the CBE cyclic cosmological model, the density perturbations arise quite diffferently
from in inflationary cosmology because (i) they originate not during expansion but during
contraction and (ii) they arise from quantum fluctuations not of an inflaton field but of the
electromagnetic field. 

\bigskip

\noindent
This keeps alive the beautiful idea that the large scale structure in the present
universe originates from quantum fluctuations in the very early universe.

\newpage

\section{Discussion}
Previously it was shown how a cyclic universe with the CBE assumption, that
the contracting universe is an introverse empty of matter, can explain the
observed flatness of the present universe without an inflationary era. 
It also led to an estimate of the time until the turnaround from expansion
to contraction of $1.3$ Ty, or about one hundred times the present age.

\bigskip

\noindent
In the present article we have analyzed the appearance of density
perturbations in the CBE model. Also, in Eq.(\ref{test}), there was an
observational method to falsify CBE.

\bigskip

\noindent
So where do we stand on the key question of whether the present expansion
era began with a big bang or a bounce at a time $t_0$ in the past? 
At least, we know $t_0 = 13.80\pm0.04$Gy. The two alternatives
coincide after the first picosecond, $10^{-12}$s, but differ completely in
the earlier universe.

\bigskip

\noindent
It is necessary to be clear on what "big bang" theory means. We take it to
mean that time starts $t_0$ ago and that at that time the density and temperature
were both extremely large. They are not necessarily infinite, as suggested by
the classical Friedmann-LeMa\^itre equation\cite{Friedmann,Lemaitre} because, for times
$t\lesssim t_{Planck}=10^{-44}$s, time itself becomes ill-defined due to quantum
fluctuations of spacetime. In the distant future, the big-bang model is normally taken
to imply expansion for an infinite time.

\bigskip

\noindent
In big-bang theory, to explain the flatness and horizon properties a brief period
of superluminal accelerates expansion helps during the extremely early universe.
Even with such inflation, however, there still remains a mystery about the 
initial conditions, especially why the entropy is so singularly low \cite{Penrose}.

\bigskip

\noindent
By ''bounce" which is our clear preference, it is meant that at the bounce time
$t_B$ satisfying
\begin{equation}
10^{-38}s <  t < 10^{-18}s
\end{equation}
the present expansion era began immediately preceded by contraction. There
was no inflation era. Nevertheless, the flatness property is
predicted and, as we have shown in this article, so are scale invariant
density perturbations at lowest order. One expects the reddening from spectral
index $n_s=1$ to the observed $n_s\simeq 0.96$ to be calculable in higher
orders, similarly to what happens in inflation.

\bigskip

\noindent
It should be added that the CBE assumption first introduced in \cite{BF},
and considerable refined here, is very speculative but, to our knowledge,
there is no alternative resolution of the Tolman conundrum\cite{Tolman}.

\bigskip

\noindent
One outstanding issue is how the turnaround and bounce occur dynamically.
If we may close with a speculation, the correct theory of quantum gravity
could have a classical limit agreeing with general relativity everywhere, except
in the close vicinity of the turnaround or bounce.


\newpage


\bigskip
\bigskip

\begin{thebibliography}{100}

\bibitem{PHF1}
P.H. Frampton, 
Int. J. Mod. Phys. {\bf A30,} 1550129 (2015).\\
{\tt arXiv:1501.03054[gr-qc] }.
\bibitem{PHF2}
P.H. Frampton. {\tt arXiv:1503.03121[gr-qc]}
\bibitem{Tolman}
R.C. Tolman, Phys. Rev. {\bf 38,} 1758 (1931).
\bibitem{Harrison}
E.R. Harrison, Phys. Rev. {\bf D1,} 2726 (1970).
\bibitem{Zeldovich}
R.A. Sunyaev and Ya.B. Zel'dovich, Astrophys. Space. Sci.  {\bf 7,} 20 (1970).
\bibitem{Peebles}
P.J.E. Peebles and J.T. Yu, Astrophy. J. {\bf 162,} 815 (1970).
\bibitem{Planck1}
P.A.R. Ade, {\it et al.} (Planck Collaboration)
{\tt arXiv:1502.01589[astro-ph.CO]}.
\bibitem{Planck2}
P.A.R. Ade, {\it et al.} (Planck Collaboration)
{\tt arXiv:1502.01592[astro-ph.CO]}.
\bibitem{Perlmutter}
S. Perlmutter, {\it et al.} (Supernova Cosmology Project).
Astrophys. J. {\bf 517,} 565 (1999).
{\tt astro-ph/9812133}.
\bibitem{Riess}
A.G. Riess, {\it at al.} (High-Z Supernova Search Team).
Astron. J. {\bf 116,} 1009 (1998).
{\tt astro-ph/9805201}.
\bibitem{BF}
L. Baum and P.H. Frampton, Phys. Rev. Lett. {\bf 98,}  071301 (2007).\\
{\tt hep-th/0610213}
\bibitem{Lyth}
D.H. Lyth and A.R. Liddle, {\it The Primordial Density Perturbation}.\\
Cambridge University Press (2009). Chap. 24.
\bibitem{Guth}
A.H. Guth, Phys. Rev. {\bf D23,} 347 (1981).
\bibitem{Linde}
A.D. Linde, Phys. Lett. {\bf B108,} 389 (1982).
\bibitem{Steinhardt}
A. Albrecht and P.J. Steinhardt, Phys. Rev. Lett. {\bf 48,} 1220 (1982).
\bibitem{Friedmann}
A. Friedmann, Z. Phys. {\bf 10,} 377 (1922).
\bibitem{Lemaitre}
G. LeMa\^itre, Ann. Sci. Soc. Brussels, {\bf A47,} 49 (1927).\\
Translation: MNRAS, {\bf 91,} 483 (1931).
\bibitem{Penrose}
R.Penrose, {\it The Road to Reality}. Vintage Books. (2005).
\end{thebibliography}
\end{document}